\documentclass[10pt,reqno]{amsart}
\usepackage[latin9]{inputenc}
\usepackage{amsmath}
\usepackage{amssymb}
\usepackage{babel}
\usepackage{comment}
\usepackage{amssymb,mathrsfs,color}
\usepackage{pinlabel}
\usepackage{amsmath, amsfonts, amsthm, verbatim, amssymb}
\usepackage{epstopdf, mathtools}
\usepackage{cite}
\usepackage{fullpage}
\usepackage{microtype}
\newcommand{\bs}{\boldsymbol}

\newcommand{\bbF}{\mathbb F}
\newcommand{\bbP}{\mathbb P}
\newcommand{\bbA}{\mathbb A}

\newcommand{\p}{\partial}
\newcommand{\bb}{\mathbb}
\allowdisplaybreaks[3]
\numberwithin{equation}{section}

\begin{document}

\title{On the synthesis of complete two-dimensional second-gradient continua: tri-pantographic fabrics}
\author{C. Rodriguez, E. Barchiesi,  S. R. Eugster, I. Giorgio, F. dell'Isola}

\begin{abstract}
We introduce a notion of completeness for two-dimensional second-gradient elastic continua and propose a microstructural route toward its synthesis. A continuum is said to be complete if the Hessian of the stored energy with respect to the second-gradient variable is locally positive definite, so that every nonzero admissible increment of the placement second gradient is quadratically controlled in the highest-order part of the energy about each configuration. Starting from the Principle of Virtual Work, we derive the constitutive relations, equilibrium equations, and admissible boundary interactions for a broad class of fibrous second-gradient continua whose stored energies depend on fiber stretch, stretch gradients, and curvature. The theory is then applied to several continua motivated by pantographic microstructures. Classical pantographic sheets are shown to be incomplete, while bi-pantographic fabrics enlarge the class of components of the second gradient detected by the energy but remain incomplete. Finally, we formulate a tri-pantographic continuum associated with a proposed three-family architecture and prove that it is complete. The examples illustrate how microstructural architecture influences both the completeness properties of an effective continuum and the pointwise form of its higher-order boundary interactions.
\end{abstract}

\maketitle

\section{Introduction}

\subsection{Synthesis of mechanical metamaterials} A central objective in the design of mechanical metamaterials is the \textit{synthesis} of architectures capable of large elastic deformations. Pantographic lattices and fabrics provide notable examples, accommodating large reversible changes of shape through bending, stretching, and rotations within their microstructure. Their effective descriptions are two-dimensional\footnote{Throughout this work, by a two-dimensional continuum we mean a continuum whose reference and current configurations are both contained in a plane.} second-gradient elastic continua, whose stored energies depend on the deformation gradient and its gradient. This enriched kinematic structure admits a broader class of boundary interactions and raises the natural question of whether the stored energy provides strictly positive quadratic control of every admissible second-gradient increment. Motivated by this question, we call a continuum \textit{complete} if, at each material point and for each fixed deformation gradient, the stored-energy Hessian with respect to the second-gradient variable is locally positive definite.

If the stored energy does not depend on particular components of the second gradient, the corresponding pointwise higher-order boundary interactions may be restricted. More generally, loss of positive definiteness renders the principal quadratic form degenerate, thereby complicating questions of well-posedness and, consequently, predictability. Although a complete one-dimensional pantographic continuum was obtained in \cite{Barchiesi2019}, its two-dimensional counterpart has remained unresolved. This leads to the central constructive question of the paper: is there a concrete candidate pantographic architecture that yields a complete two-dimensional continuum?

The word \textit{synthesis} is used in the sense of \cite{Alibert2003}, and is opposed to \textit{analysis} as an inverse problem is opposed to the corresponding direct one. In analysis, a microstructure is prescribed, for example, a truss, lattice, or network of beams governed by standard first-gradient constitutive laws, and, under an assigned scaling and suitable compactness assumptions, one seeks its effective continuum description as the cell size tends to zero. The existence and identification of such a limit depend on the hypotheses and convergence framework adopted. In synthesis, by contrast, the macroscopic theory is prescribed, and one must exhibit a family of mechanical systems having that effective limit, or prove that none exists. This program took its modern form in \cite{Alibert2003}, where a modular truss beam, now called a \textit{pantographic beam}, was constructed whose homogenized energy depends on the macroscopic second gradient and whose first-gradient contribution can be made asymptotically negligible. A deliberate separation of stiffness scales makes the arrangement comparatively soft with respect to microscopic motions associated with the macroscopic first gradient and stiff with respect to those associated with the second gradient. The construction shows that a higher-gradient effective theory can arise when the microstructure suppresses the leading Cauchy contribution. Within the variational setting studied in \cite{CamarEddine2003}, the closure of classical elastic functionals was subsequently shown to contain second- and higher-gradient energies, establishing that such targets form a meaningful class of synthesis problems. These constructions are reviewed in \cite{Barchiesi2019a}, their inverse-problem character is analyzed in \cite{Fedele2024}, and the design-first, homogenize-afterward viewpoint is termed \textit{mathematically-driven design} in \cite{dellIsola2019}.

In two spatial dimensions, the basic \textit{pantographic sheet} consists of two orthogonal families of continuous extensible Euler beams connected by ideal pivots \cite{dellIsola2016c,dellIsola2016b}. Its homogenized energy depends on the geodesic bending of the material lines, and the resulting response has been extensively studied numerically and experimentally \cite{Turco2016,Placidi2017a,Boutin2017a,dellIsola2019b,Maurin2019}. Replacing the Euler beams with pantographic beams introduces a third length scale and yields the \textit{bi-pantographic fabric}, whose energy also depends on gradients of fiber elongation \cite{Barchiesi2020a,Barchiesi2020c,Barchiesi2020b}. For the two orthogonal fiber directions, the pantographic-sheet energy detects only selected projections of the directional second derivatives, while the bi-pantographic energy controls the full directional second derivatives taken twice along each fiber family. Mixed second derivatives involving the two directions nevertheless remain undetected. Introducing a third, oblique fiber family causes its directional second derivative to couple the two pure derivatives to the missing mixed derivative. The three directional contractions can then determine every admissible second-gradient increment. This observation motivates the tri-pantographic continuum introduced in this work and identifies the third family as the minimal natural architectural addition.
	
To formulate these continua and connect their energetic properties to mechanical consequences, we employ the Principle of Virtual Work. The method goes back to Germain \cite{Germain1973b}, whose second-gradient memoir is available in English translation \cite{Germain2020} together with a commentary on its assumptions \cite{Epstein2020}, and has since been systematized in \cite{dellIsola2015,dellIsola2016,dellIsola2020c,Seppecher2020a,RodriguezDellIsola2026}. Once the internal virtual work is prescribed, the equilibrium equations, Piola stress, double stress, and associated pointwise boundary interactions follow together: forces and double forces on smooth edges, and concentrated forces at corners. The same content may be expressed in convected coordinates and in Lagrangian or Eulerian form \cite{Fedele2025}, or embedded in micromorphic theories with corresponding energy theorems \cite{Desmorat2025}. If the stored energy is independent of a specified second-gradient component, its contribution to the double stress is absent and the corresponding pointwise boundary interactions may be restricted. This local statement does not by itself determine whether independently prescribed boundary data can be realized globally. Virtual work nevertheless makes the higher-order boundary interactions explicit and thereby connects the energetic question of completeness to its mechanical consequences.
	
Related developments at the third-gradient level reinforce the same connection between architecture, energetic dependence, and boundary interactions. Planar beams governed by the third gradient of the placement, equivalently by the gradient of curvature, have been targeted in \cite{dellIsola2024} and surveyed in \cite{MurciaTerranova2025}. The zigzagged articulated parallelogram with articulated braces (ZAPAB) provides a candidate discrete route: its kinematics are established in \cite{Moschini2025}, its double-bending stiffness is identified computationally in \cite{Terranova2025}, and curvature-gradient beams and candidate truss architectures are examined in \cite{dellIsola2026}. Complementary continuum formulations derive the nonstandard equilibrium equations, work-conjugate variables, and deformation-dependent coupling of generalized external actions \cite{Fedele2022b,Fedele2022a}. These developments clarify the present second-gradient objective: as the gradient order increases, the pointwise boundary interactions become richer, making it still more important to determine whether the energy controls every admissible increment of its highest-gradient variable.
	
A proposed synthesis must also survive the passage from an ideal drawing to a specimen. First, fabrication can alter the geometry sufficiently to change the effective response. Additive manufacturing introduces substructure through infill density, deposition pattern, and polymer response \cite{Yildizdag2019,Abali2021,Aydin2022,Aydin2022a,Afshar2023}, while asymptotic homogenization connects the printed cell to generalized constitutive parameters and size effects that a Cauchy model cannot reproduce \cite{Abali2022}. The response can be highly sensitive to cell geometry \cite{Turco2019}: small nodal displacements may alter it disproportionately \cite{Turco2026}, and experiments on metallic specimens confirm the dominant influence of geometry through the onset of damage \cite{DeAngelo2019}. This sensitivity has been exploited to obtain enriched buckling \cite{Eremeyev2020}, multistability \cite{Turco2024}, and prescribed wave transmission and reflection \cite{Yildizdag2023}. Granular micromechanics has likewise predicted chiral responses realized in printed specimens \cite{Misra2020}, guided anisotropy optimization \cite{Placidi2026}, and, together with discrete-chain analyses, produced tunable frequency band gaps \cite{Nejadsadeghi2019,Placidi2024}. These results support a second practical conclusion: the sufficiency of an architecture must be established through its effective energy rather than inferred from geometry alone.

The kinematic descriptors entering the effective energy must therefore be chosen to represent the mechanisms relevant at the scale of interest. Shear compliance at nonideal lattice interconnections can symmetrize an anisotropic response \cite{desmorat2020mms,Spagnuolo2022} and, under consistent rescaling, appears as a mesoshear contribution that the effective theory must retain \cite{Spagnuolo2025}. Geometry can also be prescribed to obtain a target spatial response, as in an annular lattice woven from logarithmic spirals \cite{Ciallella2023}, while hierarchical natural materials such as bone suggest further architectures \cite{Giorgio2020a}. A coarser descriptor set may suffice for a restricted response, as in an equivalent Timoshenko-like beam \cite{Pancella2025}; other settings require an enlarged description, such as a space-time representative volume for dissipative systems \cite{Chekeres2024} or active microelectromechanical textures for exoskeletons and soft robotics \cite{Carcaterra2026}. In the present work, the descriptors are embedded fibers' stretch, stretch gradients, and curvature. Completeness asks whether an energy constructed from them gives positive quadratic control of every admissible second-gradient increment.
	
Against this background, we propose completeness as a design criterion connecting the energetic control of the second gradient to the higher-order response of a continuum. The central result is the construction of a complete tri-pantographic continuum obtained by adding an oblique third fiber family to a bi-pantographic fabric. A comparison with classical pantographic sheets and bi-pantographic fabrics reveals the distinct structural sources of their incompleteness and shows how the third family removes the remaining energetic degeneracy. The analysis is carried out within a general variational framework for fibrous second-gradient continua, which also identifies the associated pointwise higher-order boundary interactions. A corresponding discrete three-family construction is proposed as a concrete candidate architecture, although its rigorous homogenization remains open.

\subsection{Overview}

Section 2 develops the kinematics of two-dimensional continua containing embedded fiber families, including the relevant expressions for fiber stretch, stretch gradients, curvature, and their variations. Section 3 formulates the variational theory for a broad class of fibrous second-gradient elastic continua. It derives the constitutive relations, equilibrium equations, and pointwise boundary interactions, and then introduces completeness and characterizes it through local positive definiteness of the stored-energy Hessian with respect to the second-gradient variable.
	
Section 4 applies this criterion to continua motivated by pantographic microstructures. It identifies the respective sources of incompleteness in classical pantographic sheets and bi-pantographic fabrics, constructs the complete tri-pantographic continuum, and presents the corresponding discrete three-family architecture. Section 5 summarizes the results and discusses possible connections among completeness, higher-order boundary interactions, and the well-posedness of the associated boundary value problems.

\section{Kinematics}

In this section, we develop the kinematics of continua containing embedded fiber families. Particular attention is devoted to the geometric quantities associated with the fibers, including stretch, stretch gradients, and curvature, since these variables determine the second-gradient stored energies studied later in the paper.

\subsection{Preliminaries}

For the modeling of two-dimensional continua, we work in a {two-dimensional} Euclidean space identified with $\mathbb{R}^2$ after fixing an origin and an orthonormal basis for the translation space. The reference configuration of the body is the closure of a domain $\Omega \subset \mathbb{R}^2$ with topological boundary $\partial \Omega$, assumed to be Lipschitz and piecewise smooth. The finitely many points at which differentiability fails are called \textit{corners}, and we denote their collection by $\partial \partial \Omega$. The connected components of $\partial \Omega \setminus \partial \partial \Omega$ are referred to as the \textit{edges} of the boundary. Each edge is oriented consistently with its outward pointing unit normal field $\boldsymbol N \in \mathbb{R}^2$.

The placement $\boldsymbol \chi \colon \Omega \to \mathbb{R}^2$ maps points $\boldsymbol X \in \Omega$ in the reference configuration to spatial points $\boldsymbol x \in \omega := \boldsymbol \chi(\Omega)$ in the current configuration. To distinguish reference and spatial quantities, we introduce orthonormal, positively oriented bases $\{\boldsymbol E_1, \boldsymbol E_2\}$ and $\{\boldsymbol e_1, \boldsymbol e_2\}$ for the reference and current configurations, respectively. With this convention, the placement map is expressed as
\begin{equation}
	\boldsymbol x = x_i \boldsymbol e_i = \boldsymbol \chi(\boldsymbol X) = \chi_i(X_A \boldsymbol E_A)\boldsymbol e_i,
\end{equation}
where we adopt Einstein's summation convention and all Latin indices take values in $\{1,2\}$. Uppercase indices refer to components with respect to $\{\boldsymbol E_1, \boldsymbol E_2\}$, while lowercase indices refer to $\{\boldsymbol e_1, \boldsymbol e_2\}$.

The first and second gradients of the placement, $\boldsymbol F = \nabla \boldsymbol \chi$ and $\bbF = \nabla \boldsymbol F$, have components
\begin{equation}\label{eq:first_second_placement_gradient}
	F_{iA} = \frac{\partial \chi_i}{\partial X_A}, \quad
	\bbF_{iAB} = \frac{\partial^2 \chi_i}{\partial X_A \partial X_B},
\end{equation}
both defined as functions of $\boldsymbol X$.

To define variations, let $\hat{\boldsymbol \chi} \colon \mathbb{R} \times \Omega \to \mathbb{R}^2$ be a {one-parameter} family of placements such that $\boldsymbol \chi(\boldsymbol X) = \hat{\boldsymbol \chi}(0,\boldsymbol X)$. The associated variation of the placement is given by
\begin{equation}
	\delta \boldsymbol \chi(\boldsymbol X) := \frac{\partial \hat{\boldsymbol \chi}}{\partial \varepsilon}(0,\boldsymbol X) = \frac{\partial \hat \chi_i}{\partial \varepsilon}(0,\boldsymbol X)\boldsymbol e_i,
\end{equation}
which defines {the \textit{virtual displacement field}}. The induced variations of the first and second gradients follow by differentiation and commutation of partial derivatives:
\begin{align}
	\delta F_{iA} &= \frac{\partial}{\partial \varepsilon}\Bigl (\frac{\partial \hat \chi_i}{\partial X_A}\Bigr )\Big |_{\varepsilon=0}
	= \frac{\partial \delta \chi_i}{\partial X_A}
	= (\nabla \delta \boldsymbol \chi)_{iA}, \\
	\delta \bbF_{iAB} &= \frac{\partial}{\partial \varepsilon}\Bigl (\frac{\partial^2 \hat \chi_i}{\partial X_A \partial X_B}\Bigr )\Big |_{\varepsilon=0}
	= \frac{\partial^2 \delta \chi_i}{\partial X_A \partial X_B}
	= (\nabla \nabla \delta \boldsymbol \chi)_{iAB}.
\end{align}

\subsection{Fiber kinematics}

An embedded fiber in the reference configuration $\Omega \subset \mathbb{R}^2$ is described by a smooth curve $\boldsymbol \Gamma = \boldsymbol \Gamma(s)$, parameterized by arc length $s$. Under the placement map $\boldsymbol \chi \colon \Omega \to \mathbb{R}^2$, the fiber is mapped to the convected curve
\[
\boldsymbol \gamma(s) = \boldsymbol \chi(\boldsymbol \Gamma(s))
\]
in the current configuration.

Denoting differentiation with respect to $s$ by a prime, the unit tangent vector $\boldsymbol D = \boldsymbol \Gamma'$ is mapped to the convected tangent vector
\begin{equation}\label{eq:tangent_vector_mapping}
	\boldsymbol d = \boldsymbol \gamma' = (\boldsymbol \chi \circ \boldsymbol \Gamma)' 
	= (\nabla \boldsymbol \chi \circ \boldsymbol \Gamma)\boldsymbol \Gamma'
	= (\boldsymbol F \circ \boldsymbol \Gamma)\boldsymbol D.
\end{equation}
Writing $\boldsymbol D = D_A \boldsymbol E_A$ and $\boldsymbol d = d_i \boldsymbol e_i$, this relation takes the form
\begin{equation}\label{eq:current_tangent_vector_index}
	d_i = F_{iA} D_A.
\end{equation}

The convected tangent vector admits the representation
\begin{equation}\label{eq:current_tangent_vector}
	\boldsymbol d = \rho \boldsymbol \tau 
	= \rho \big( \cos \vartheta \boldsymbol e_1 + \sin \vartheta \boldsymbol e_2 \big)  ,
\end{equation}
where
\begin{equation}\label{eq:fiber_stretch_and_angle}
	\rho := |\boldsymbol d| = (\boldsymbol d \cdot \boldsymbol d)^{1/2}  , 
	\quad
	\vartheta := \arctan\!\left( \frac{\boldsymbol d \cdot \boldsymbol e_2}{\boldsymbol d \cdot \boldsymbol e_1} \right)
\end{equation}
denote the fiber \textit{stretch} and \textit{inclination angle}, respectively. Here, the unit tangent vector is given by
\begin{equation}\label{eq:tangent_vector_curve}
	\boldsymbol \tau = \cos \vartheta \boldsymbol e_1 + \sin \vartheta \boldsymbol e_2  .
\end{equation}
Let $\boldsymbol R = \boldsymbol e_2 \otimes \boldsymbol e_1 - \boldsymbol e_1 \otimes \boldsymbol e_2$, which can be seen as a rotation around $90^\circ$ in the counterclockwise direction. A vector perpendicular to $\boldsymbol \tau$ is given by
\begin{equation}\label{eq:tangent_perp_vector_curve}
	\boldsymbol \nu = \boldsymbol R \boldsymbol \tau = -\sin{\vartheta} \boldsymbol e_1 + {\cos{\vartheta}\boldsymbol e_2}   .
\end{equation}
The second derivative of the curve $\boldsymbol \gamma$ with respect to reference arc length is
\begin{equation}
	\boldsymbol g = \boldsymbol \gamma'' = \boldsymbol d' = \rho' \boldsymbol \tau + \rho \boldsymbol \tau' = \rho' \boldsymbol \tau + \rho \vartheta' \boldsymbol \nu  ,
\end{equation}
where we have used the identity
\begin{equation}
	\boldsymbol \tau' = \vartheta' \boldsymbol \nu  ,
\end{equation}
which follows readily from \eqref{eq:tangent_vector_curve} and \eqref{eq:tangent_perp_vector_curve}. We call $\rho'$ the \textit{stretch gradient} and $\vartheta'$ the \textit{curvature}. 	In the differential geometry of planar curves, curvature is usually defined as the rate of change of the inclination angle $\vartheta$ with respect to arc length of the current configuration $\boldsymbol \gamma$, i.e., $\rho^{-1}\vartheta'$.

If a continuous distribution of fibers is embedded in the body, the kinematical quantities introduced above can be viewed as fields defined on $\Omega$. In particular, a fiber curve passes through each point $\boldsymbol X \in \Omega$. In what follows, we restrict attention to continua containing initially straight and parallel fibers. We allow for multiple fiber families, each characterized by a constant unit direction field
\[
\boldsymbol D_\alpha = D^{(\alpha)}_A \boldsymbol E_A  .
\]
For each family $\alpha$ and each point $\boldsymbol X \in \Omega$, there exists a curve $\boldsymbol \Gamma_\alpha$ such that
\begin{align}
\boldsymbol \Gamma_\alpha'(s) = \boldsymbol D_\alpha, \quad \boldsymbol \Gamma_\alpha(s) = \boldsymbol X. \label{eq:gammaX}
\end{align}

Let $f$ be a tensor valued function defined on $\Omega$, $\boldsymbol X \in \Omega$, and $\boldsymbol \Gamma_\alpha$ be as in \eqref{eq:gammaX}. The derivative of $f$ at $\boldsymbol X$ along the fibers of the $\alpha$ family is defined by
\begin{equation}
	f_{,\alpha}(\boldsymbol X) := (f \circ \boldsymbol \Gamma_\alpha)'(s) 
	= (\nabla f  \boldsymbol D_\alpha)(\boldsymbol X).
\end{equation}
{Equivalently, the preceding relation may be written as}
\begin{align}
	f_{,\alpha} = D^{(\alpha)}_{B} \frac{\partial}{\partial X_B} f. \label{eq:fiberderivative}
\end{align}

Unless stated otherwise, repeated appearance of Greek letters does not imply summation. In view of \eqref{eq:tangent_vector_mapping} and \eqref{eq:fiber_stretch_and_angle}, the stretch of fibers in the $\alpha$ family is given by
\begin{equation}\label{eq:alpha_fiber_stretch}
	\rho_\alpha = |\boldsymbol d_\alpha| 
	= \big[(\boldsymbol F \boldsymbol D_\alpha) \cdot (\boldsymbol F \boldsymbol D_\alpha)\big]^{1/2}
	= \big[F_{iA} D^{(\alpha)}_A F_{iB} D^{(\alpha)}_B\big]^{1/2}  ,
\end{equation}
while the inclination angle is
\begin{equation}
	\vartheta_\alpha 
	= \arctan\!\left(\frac{\boldsymbol d_\alpha \cdot \boldsymbol e_2}{\boldsymbol d_\alpha \cdot \boldsymbol e_1}\right)
	= \arctan\!\left(\frac{F_{2A} D^{(\alpha)}_A}{F_{1B} D^{(\alpha)}_B}\right)  .
\end{equation}
The associated unit tangent and normal vectors to the convected fibers are therefore
\begin{equation}\label{eq:current_unit_tangent_vector}
	\boldsymbol \tau_\alpha 
	= \rho_\alpha^{-1} \boldsymbol d_\alpha 
	= \rho_\alpha^{-1} F_{iA} D^{(\alpha)}_A \boldsymbol e_i  , 
	\qquad 
	\boldsymbol \nu_\alpha = \boldsymbol R \boldsymbol \tau_\alpha  .
\end{equation}

The virtual displacement induces variations of the fiber stretch and inclination angle. From \eqref{eq:alpha_fiber_stretch}, we obtain
\begin{equation}\label{eq:variation_fiber_stretch}
	\delta \rho_\alpha
	= \rho_\alpha^{-1}\delta F_{iA}D^{(\alpha)}_A F_{iB}D^{(\alpha)}_B
	= \delta F_{iA}\tau^{(\alpha)}_i D^{(\alpha)}_A
	= \delta \boldsymbol F \cdot (\boldsymbol \tau_\alpha \otimes \boldsymbol D_\alpha),
\end{equation}
where $\cdot$ denotes contraction over all indices. 
The derivative of the fiber stretch along the fibers of the $\alpha$ family is similarly determined by
\begin{equation}\label{eq:fiber_stretch_gradient}
	\rho_{\alpha,\alpha}
	= \rho_\alpha^{-1}F_{iA,B}D^{(\alpha)}_A F_{iC}D^{(\alpha)}_C D^{(\alpha)}_B
	= \bbF_{iAB}\tau^{(\alpha)}_i D^{(\alpha)}_A D^{(\alpha)}_B
	= \bbF \cdot (\boldsymbol \tau_\alpha \otimes \boldsymbol D_\alpha \otimes \boldsymbol D_\alpha).
\end{equation}

Recalling \eqref{eq:current_unit_tangent_vector}, the variation of the inclination angle is
\begin{equation}\label{eq:variation_inclination_angle}
	\begin{aligned}
		\delta \vartheta_\alpha
		&= \rho_\alpha^{-2}
		\left[
		(\delta \boldsymbol d_\alpha \cdot \boldsymbol e_2)(\boldsymbol d_\alpha \cdot \boldsymbol e_1)
		-(\boldsymbol d_\alpha \cdot \boldsymbol e_2)(\delta \boldsymbol d_\alpha \cdot \boldsymbol e_1)
		\right] \\
		&= \rho_\alpha^{-2}\delta \boldsymbol d_\alpha \cdot \boldsymbol R\boldsymbol d_\alpha \\
		&= \rho_\alpha^{-1}\delta \boldsymbol d_\alpha \cdot \boldsymbol \nu_\alpha \\
		&= \rho_\alpha^{-1}\nu^{(\alpha)}_i\delta F_{iA}D^{(\alpha)}_A
		= \delta \boldsymbol F \cdot \rho_\alpha^{-1}\boldsymbol \nu_\alpha \otimes \boldsymbol D_\alpha .
	\end{aligned}
\end{equation}

The fiber derivative of the inclination angle, which we call the fiber curvature, is
\begin{equation}\label{eq:fiber_curvature}
	\begin{aligned}
		\vartheta_{\alpha,\alpha}
		&= \rho_\alpha^{-2}
		\left[
		(\boldsymbol d_{\alpha,\alpha} \cdot \boldsymbol e_2)(\boldsymbol d_\alpha \cdot \boldsymbol e_1)
		-(\boldsymbol d_\alpha \cdot \boldsymbol e_2)(\boldsymbol d_{\alpha,\alpha} \cdot \boldsymbol e_1)
		\right] \\
		&= \rho_\alpha^{-2}\boldsymbol d_{\alpha,\alpha} \cdot \boldsymbol R\boldsymbol d_\alpha
		= \rho_\alpha^{-1}\boldsymbol d_{\alpha,\alpha} \cdot \boldsymbol \nu_\alpha \\
		&= \rho_\alpha^{-1}\nu^{(\alpha)}_iF_{iA,B}D^{(\alpha)}_A D^{(\alpha)}_B \\
		&= \rho_\alpha^{-1}\nu^{(\alpha)}_i\bbF_{iAB}D^{(\alpha)}_A D^{(\alpha)}_B \\
		&= \bbF \cdot (
		\rho_\alpha^{-1}\boldsymbol \nu_\alpha \otimes \boldsymbol D_\alpha \otimes \boldsymbol D_\alpha).
	\end{aligned}
\end{equation}

Since
\[
\boldsymbol \tau_\alpha
= \cos \vartheta_\alpha \boldsymbol e_1
+ \sin \vartheta_\alpha \boldsymbol e_2,
\qquad
\boldsymbol \nu_\alpha
= -\sin \vartheta_\alpha \boldsymbol e_1
+ \cos \vartheta_\alpha \boldsymbol e_2,
\]
their variations are
\begin{equation}\label{eq:variation_tangent_vector}
	\delta \boldsymbol \tau_\alpha
	= \delta \vartheta_\alpha \boldsymbol \nu_\alpha
\end{equation}
and
\begin{equation}\label{eq:variation_perpendicular}
	\delta \boldsymbol \nu_\alpha
	= -\delta \vartheta_\alpha \boldsymbol \tau_\alpha .
\end{equation}
The corresponding fiber derivatives are
\begin{equation}\label{eq:tangent_vector_fiber_derivative}
	\boldsymbol \tau_{\alpha,\alpha}
	= \vartheta_{\alpha,\alpha}\boldsymbol \nu_\alpha,
	\qquad
	\boldsymbol \nu_{\alpha,\alpha}
	= -\vartheta_{\alpha,\alpha}\boldsymbol \tau_\alpha .
\end{equation}

We next compute the variation of the fiber stretch gradient. Using
\eqref{eq:fiber_stretch_gradient}, \eqref{eq:variation_tangent_vector},
and \eqref{eq:fiber_curvature}, we find
\begin{equation}\label{eq:variation_fiber_stretch_gradient}
	\begin{aligned}
		\delta \rho_{\alpha,\alpha}
		&=
		\delta \bbF_{iAB}\tau^{(\alpha)}_iD^{(\alpha)}_A D^{(\alpha)}_B
		+\bbF_{iAB}\delta \tau^{(\alpha)}_iD^{(\alpha)}_A D^{(\alpha)}_B \\
		&=
		\delta \bbF_{iAB}\tau^{(\alpha)}_iD^{(\alpha)}_A D^{(\alpha)}_B
		+\delta \vartheta_\alpha \nu^{(\alpha)}_i\bbF_{iAB}D^{(\alpha)}_A D^{(\alpha)}_B \\
		&=
		\delta \bbF_{iAB}\tau^{(\alpha)}_iD^{(\alpha)}_A D^{(\alpha)}_B
		+\delta F_{iA}\nu^{(\alpha)}_iD^{(\alpha)}_A\vartheta_{\alpha,\alpha} \\
		&=
		\delta \bbF \cdot
		(\boldsymbol \tau_\alpha \otimes \boldsymbol D_\alpha \otimes \boldsymbol D_\alpha)
		+
		\delta \boldsymbol F \cdot
		(\vartheta_{\alpha,\alpha}\boldsymbol \nu_\alpha \otimes \boldsymbol D_\alpha).
	\end{aligned}
\end{equation}

Finally, using \eqref{eq:variation_fiber_stretch}, \eqref{eq:variation_perpendicular},
\eqref{eq:variation_inclination_angle}, and \eqref{eq:fiber_stretch_gradient},
the variation of the fiber curvature is
\begin{equation}\label{eq:variation_fiber_curvature}
	\begin{aligned}
		\delta \vartheta_{\alpha,\alpha}
		&=
		-\rho_\alpha^{-2}\delta \rho_\alpha
		\bbF_{iAB}D^{(\alpha)}_A D^{(\alpha)}_B\nu^{(\alpha)}_i
		+\rho_\alpha^{-1}\delta \bbF_{iAB}D^{(\alpha)}_A D^{(\alpha)}_B\nu^{(\alpha)}_i \\
		&\qquad
		+\rho_\alpha^{-1}\bbF_{iAB}D^{(\alpha)}_A D^{(\alpha)}_B
		\delta \nu^{(\alpha)}_i \\
		&=
		-\rho_\alpha^{-1}\delta F_{iA}\tau_i^{(\alpha)}D^{(\alpha)}_A\vartheta_{\alpha,\alpha}
		+\rho_\alpha^{-1}\delta \bbF_{iAB}D^{(\alpha)}_A D^{(\alpha)}_B\nu^{(\alpha)}_i \\
		&\qquad
		-\rho_\alpha^{-2}\rho_{\alpha,\alpha}\nu^{(\alpha)}_i\delta F_{iA}D^{(\alpha)}_A \\
		&=
		\delta \bbF_{iAB}
		\left(\rho_\alpha^{-1}\nu^{(\alpha)}_iD^{(\alpha)}_A D^{(\alpha)}_B\right) \\
		&\qquad
		+\delta F_{iA}
		\left(
		-\rho_\alpha^{-1}\vartheta_{\alpha,\alpha}\tau_i^{(\alpha)}D^{(\alpha)}_A
		-\rho_\alpha^{-2}\rho_{\alpha,\alpha}\nu^{(\alpha)}_iD^{(\alpha)}_A
		\right) \\
		&=
		\delta \bbF \cdot
		\rho_\alpha^{-1}\boldsymbol \nu_\alpha \otimes \boldsymbol D_\alpha \otimes \boldsymbol D_\alpha \\
		&\qquad
		+
		\delta \boldsymbol F \cdot
		\left(
		-\rho_\alpha^{-1}\vartheta_{\alpha,\alpha}\boldsymbol \tau_\alpha \otimes \boldsymbol D_\alpha
		-\rho_\alpha^{-2}\rho_{\alpha,\alpha}\boldsymbol \nu_\alpha \otimes \boldsymbol D_\alpha
		\right).
	\end{aligned}
\end{equation}

\section{Two-dimensional second-gradient continua}

In this section, we develop the variational formulation of the two-dimensional second-gradient continua studied in this work. Starting from the Principle of Virtual Work, we derive the governing equilibrium equations, admissible boundary interactions, and constitutive relations associated with stored energies depending on fiber kinematic variables.

\subsection{Virtual work functional, Piola stress and double stress} In general, a two-dimensional second-gradient continuum is a continuum whose internal virtual work functional depends on both the first and second gradients of the virtual displacement. More precisely, the internal virtual work is assumed to take the form
\begin{equation}\label{eq:virtual_work_second_gradient}
	W^\mathrm{int}(\Omega', \delta \boldsymbol \chi)
	\coloneqq
	-\int_{\Omega'}
	\big(
	\boldsymbol P \cdot \delta \boldsymbol F
	+
	\bbP \cdot \delta \bbF
	\big)
	dA
	=
	-\int_{\Omega'}
	\big(
	P_{iA}\delta F_{iA}
	+
	\bbP_{iAB}\delta \bbF_{iAB}
	\big)
	dA,
\end{equation}
where $\Omega' \subseteq \Omega$, and \(\boldsymbol P\) and \(\bbP\) denote the Piola stress and double-stress tensors, respectively.

Let \(U \subseteq \mathrm{Lin}^+\) be open, and let \(V\) denote an open subset of the space of third-order tensors symmetric in the last two indices. We consider a two-dimensional second-gradient elastic continuum with stored energy
\begin{align}
	W = W(\boldsymbol F,\bbF;\boldsymbol X),
\end{align}
where \(W\) is twice continuously differentiable on \(U\times V\times\Omega\). The associated constitutive equations are
\begin{equation}
	P_{iA}
	=
	\frac{\partial W}{\partial F_{iA}},
	\qquad
	\bbP_{iAB}
	=
	\frac{\partial W}{\partial \bbF_{iAB}}.
\end{equation}

In anticipation of applying the Principle of Virtual Work, 
\begin{align}
W^{\mathrm{int}}(\Omega', \delta \boldsymbol \chi) + W^\mathrm{ext}(\Omega', \delta \boldsymbol \chi) = 0, \label{eq:PVW}
\end{align}
for all Lipschitz, piecwise smooth $\Omega' \subseteq \Omega$, and admissible virtual displacements $\delta \boldsymbol \chi$. Repeated integration by parts shows that the class of external work functionals compatible with a second-gradient internal work functional takes the form
\begin{equation}
	W^\mathrm{ext}(\Omega',\delta\boldsymbol\chi)
	=
	\int_{\Omega'}
	\boldsymbol F^{\Omega'} \cdot \delta\boldsymbol\chi
	\, dA
	+
	\int_{\partial\Omega'}
	\left(
	\boldsymbol F^{\partial\Omega'}\cdot\delta\boldsymbol\chi
	+
	\boldsymbol D^{\partial\Omega'}\cdot
	\frac{\partial \delta\boldsymbol\chi}{\partial N}
	\right)
	dL
	+
	\sum_{i=1}^{n_\mathrm c}
	\boldsymbol F^i\cdot
	\delta\boldsymbol\chi(\boldsymbol X_i),
\end{equation}
where \(n_\mathrm c\) denotes the number of corners of the boundary $\partial \Omega'$ and
\begin{equation}
	\frac{\partial \delta\boldsymbol\chi}{\partial N}
	:=
	[\nabla(\delta\boldsymbol\chi)]\boldsymbol N
\end{equation}
is the normal derivative of the virtual displacement in the direction of the outward referential unit normal \(\boldsymbol N\). Unlike a classical first-gradient continuum, a second-gradient continuum admits additional classes of external interactions. Besides standard body and boundary forces, two-dimensional second-gradient continua can support concentrated forces acting at boundary corners and line double-forces distributed along the boundary. The line double-forces are work conjugate to the normal derivative of the virtual displacement and therefore represent higher-order contact interactions associated with the second-gradient structure of the continuum.

Defining
\begin{equation}
	\mathrm{Div}\,\bbP
	:=
	\bbP_{iAB,B}\boldsymbol e_i \otimes \boldsymbol E_A
\end{equation}
and introducing
\begin{equation}\label{eq:P_bar}
	\bar{\boldsymbol P}
	\coloneqq
	\boldsymbol P - \mathrm{Div}\, \bbP,
\end{equation}
the Principle of Virtual Work \eqref{eq:PVW} applied to $\Omega' = \Omega$ yields the following boundary value problem. The equilibrium equations are
\begin{equation}\label{eq:eqm_Lagrangian}
	\mathrm{Div}\,\bar{\boldsymbol P}
	+
	\boldsymbol F^\Omega
	=
	\boldsymbol 0
	\qquad
	\mathrm{in}\;\Omega.
\end{equation}

The associated boundary conditions along the smooth edges of the boundary are\footnote{The action of a third-order tensor $\mathbb A$ on a second-order tensor $\bs B$ is given in components via 
\begin{align}
	\mathbb A[\bs B] = \bbA_{iCD}B_{CD} \bs e_i. 
\end{align}} {Let $S$ denote arc length along each oriented edge, and let $d/dS$ denote the intrinsic tangential derivative.}
\begin{equation}\label{eq:boundary_conditions_edges_2nd_gradient}
	\begin{aligned}
		\boldsymbol F^{\partial\Omega}
		&=
		\bar{\boldsymbol P}\boldsymbol N
		-
		{\frac{d}{dS}\big(\bbP[\boldsymbol T\otimes\boldsymbol N]\big)},
		\\
		\boldsymbol D^{\partial\Omega}
		&=
		\bbP[\boldsymbol N\otimes\boldsymbol N],
	\end{aligned}
	\qquad
	\mathrm{on}\;\partial\Omega,
\end{equation}
where \(\boldsymbol T\) denotes the unit tangent vector field along the boundary. At corners of the boundary, we have
\begin{equation}\label{eq:bc_edges_Lagrangian}
	\boldsymbol F^i
	=
	(\bbP[\boldsymbol T\otimes\boldsymbol N])^-
	-
	(\bbP[\boldsymbol T\otimes\boldsymbol N])^+
	\qquad
	\mathrm{on}\;\partial\partial\Omega,
\end{equation}
where the superscripts \(+\) and \(-\) denote the one-sided limits obtained by approaching the corner along the boundary from the clockwise and counterclockwise directions, respectively.

\subsection{Definition of complete two-dimensional elastic continua} {In this work, completeness characterizes elastic second-gradient continua whose stored energies provide strictly positive quadratic control of every admissible second-gradient increment. This is stronger than mere dependence on all components of the second gradient: it is a local strong-convexity requirement with respect to the second-gradient variable. We emphasize that completeness imposes \textit{no convexity requirement on the first-gradient variable}; indeed, global convexity with respect to the deformation gradient is well known to place unphysical restrictions on admissible elastic stored energies. As shown below, the strong-convexity requirement in the second-gradient variable makes the highest-order part of the stored energy for perturbations about a given configuration positive definite in the second gradient.}

We say that a two-dimensional second-gradient elastic continuum is \textit{complete} if, for every fixed triple in the domain of the stored energy,
\[
(\boldsymbol F_0,\bbF_0,\boldsymbol X_0)
\in
U\times V\times \Omega,
\]
the mapping
\begin{align}
	\bbF
	\mapsto
	W(\boldsymbol F_0,\bbF;\boldsymbol X_0)
\end{align}
{is strongly convex on some convex neighborhood of $\bbF_0$ contained in $V$.}

Equivalently, since \(W\) is twice continuously differentiable, completeness is equivalent to the existence, for every
\[
(\boldsymbol F_0,\bbF_0,\boldsymbol X_0)
\in
U\times V\times \Omega,
\]
{of a convex neighborhood $N_{\bbF_0}\subset V$ and a constant \(c(\boldsymbol F_0,\bbF_0,\boldsymbol X_0)>0\) such that}
\begin{align}
	\bb A \cdot D^2_{\bbF\bbF}
	W(\boldsymbol F_0,\bbF;\boldsymbol X_0)
	[\bbA]
	\geq
	c(\boldsymbol F_0,\bbF_0,\boldsymbol X_0)
	|\bbA|^2 \label{eq:convcond}
\end{align}
{for all $\bbF\in N_{\bbF_0}$}
and all admissible third-order tensors \(\bbA\). In index notation using the Einstein summation convention,
\begin{align}
	\frac{\partial^2 W}
	{\partial \bbF_{iAB}\partial \bbF_{jCD}}
	(\boldsymbol F_0,\bbF;\boldsymbol X_0)
	\bbA_{iAB}\bbA_{jCD}
	\geq
	c(\boldsymbol F_0,\bbF_0,\boldsymbol X_0)
	\bbA_{iAB}\bbA_{iAB},
\end{align}
{for all $\bbF\in N_{\bbF_0}$}
and all admissible third-order tensors \(\bbA\). We note that it follows from \eqref{eq:convcond}, continuity of the second derivative of \(W\), and a standard compactness argument that, for every compact subset
\[
K \subset U \times V \times \Omega,
\]
there exists a constant \(c_K>0\) such that
\begin{align}
	\bb A \cdot D^2_{\bbF\bbF}
	W(\boldsymbol F,\bbF;\boldsymbol X)
	[\bbA]
	\geq
	c_K
	|\bbA|^2 \label{eq:convcondcomp}
\end{align}
for all \((\boldsymbol F,\bbF,\boldsymbol X)\in K\) and all admissible third-order tensors \(\bbA\). 

As an illustrative example, consider a stored energy of the form
\begin{align}
	W
	=
	W_0(\boldsymbol F;\boldsymbol X)
	+
	\frac12
	\bbF \cdot \boldsymbol{\mathsf C}(\boldsymbol F;\boldsymbol X)[\bbF], \label{eq:examplestored}
\end{align}
where \(\boldsymbol{\mathsf C}\) is a sixth-order tensor field possessing the usual major and minor symmetries.\footnote{That is, \(\boldsymbol{\mathsf C}\) is symmetric in its final two indices and symmetric under interchange of the first and second groups of three indices.} In this case, the associated second-gradient continuum is complete if and only if the quadratic form
\begin{align}
	\bbF \mapsto \bbF \cdot \boldsymbol{\mathsf C}(\boldsymbol F;\boldsymbol X)[\bbF]
\end{align}
is positive definite for every \((\boldsymbol F,\boldsymbol X)\in U\times\Omega\). In particular, completeness guarantees that \textit{every nontrivial second-order derivative contributes positively} to the stored energy \eqref{eq:examplestored}.

We next show that completeness ensures that every nontrivial second-gradient perturbation contributes to the highest-order part of the quadratic approximation of the stored energy. Let
\[
(\boldsymbol F_0,\bbF_0)
\in
U\times V
\]
and consider perturbations of the form
\begin{align}
	\boldsymbol F
	=
	\boldsymbol F_0+\nabla\boldsymbol u,
	\qquad
	\bbF
	=
	\bbF_0+\nabla\nabla\boldsymbol u.
\end{align}
The quadratic part of the Taylor expansion of \(W\) about
\((\boldsymbol F_0,\bbF_0)\)
is
\begin{align}
	Q({\nabla\boldsymbol u,\nabla\nabla\boldsymbol u}; \boldsymbol X_0) &= 
	\frac12
	\nabla \boldsymbol u \cdot D^2_{\boldsymbol F\boldsymbol F}
	W(\boldsymbol F_0,\bbF_0; \boldsymbol X_0)
	[\nabla\boldsymbol u]
	+ \nabla\boldsymbol u \cdot 
	D^2_{\bb F \boldsymbol F}
	W(\boldsymbol F_0,\bbF_0; \boldsymbol X_0)
	[\nabla\nabla\boldsymbol u] \\
	&+
	\frac12 \nabla \nabla \boldsymbol u \cdot 
	D^2_{\bbF\bbF}
	W(\boldsymbol F_0,\bbF_0; \boldsymbol X_0)
	[\nabla\nabla\boldsymbol u].
\end{align}
Since the continuum is complete,
\begin{align}
	\frac12 \nabla\nabla\boldsymbol u \cdot
	D^2_{\bbF\bbF}
	W(\boldsymbol F_0,\bbF_0, \boldsymbol X_0)
	[\nabla\nabla\boldsymbol u]
	\geq
	\frac{
		c(\boldsymbol F_0,\bbF_0, \boldsymbol X_0)
	}{2}
	|\nabla\nabla\boldsymbol u|^2.
\end{align}
Hence, completeness ensures that the highest-order part of the stored energy for perturbations about $(\boldsymbol F_0, \bb F_0)$ controls all components of the second gradient \(\nabla\nabla\boldsymbol u\).

\subsection{Stored energies depending on fiber kinematic variables}
Suppose that the stored energy depends on the kinematic variables {associated with} one or more families of fibers,
\begin{align}
	W = W(\rho_\alpha, \rho_{\alpha, \alpha}, {\vartheta_{\alpha, \alpha}}).
\end{align}
Then using the variations of the fiber stretch \eqref{eq:variation_fiber_stretch} and the fiber curvature \eqref{eq:variation_fiber_curvature} we obtain
\begin{align}
	\delta W & =  \sum_{\alpha} \bigg( 
\frac{\partial W}{\partial \rho_\alpha} \delta \rho_\alpha + \frac{\partial W}{\partial \rho_{\alpha,\alpha}} \delta \rho_{\alpha,\alpha} + 
\frac{\partial W}{\partial \vartheta_{\alpha,\alpha}} \delta \vartheta_{\alpha,\alpha} \bigg) \\
& = \sum_\alpha \bigg(\bigg[\frac{\partial W}{\partial \rho_\alpha} -\rho_\alpha^{-1} \vartheta_{\alpha,\alpha} \frac{\partial W}{\partial \vartheta_{\alpha,\alpha}} \bigg] \boldsymbol \tau_\alpha \otimes \boldsymbol D_\alpha 
+ \bigg[\frac{\partial W}{\partial \rho_{\alpha, \alpha}} {\vartheta_{\alpha, \alpha}}-\frac{\partial W}{\partial \vartheta_{\alpha,\alpha}} \rho_\alpha^{-2}\rho_{\alpha,\alpha}\bigg]\boldsymbol \nu_\alpha \otimes \boldsymbol D_\alpha \bigg) \cdot \delta \boldsymbol F  \\ 
&+ \sum_\alpha  \bigg(\frac{\partial W}{\partial \rho_{\alpha,\alpha}} \boldsymbol \tau_\alpha \otimes \boldsymbol D_\alpha \otimes \boldsymbol D_\alpha + \frac{\partial W}{\partial \vartheta_{\alpha,\alpha}}\rho_\alpha^{-1} \boldsymbol \nu_\alpha \otimes \boldsymbol D_\alpha \otimes \boldsymbol D_\alpha \bigg)\cdot \delta \bbF.
\end{align}
{By comparing the preceding expression with} the internal virtual work functional \eqref{eq:virtual_work_second_gradient}, we obtain the constitutive relations for the Piola stress and double stress 
\begin{align}
	\boldsymbol P &= \sum_\alpha \bigg(\bigg[\frac{\partial W}{\partial \rho_\alpha} -\rho_\alpha^{-1} \vartheta_{\alpha,\alpha} \frac{\partial W}{\partial \vartheta_{\alpha,\alpha}} \bigg] \boldsymbol \tau_\alpha \otimes \boldsymbol D_\alpha 
	+ \bigg[\frac{\partial W}{\partial \rho_{\alpha, \alpha}} {\vartheta_{\alpha, \alpha}}-\frac{\partial W}{\partial \vartheta_{\alpha,\alpha}} \rho_\alpha^{-2}\rho_{\alpha,\alpha}\bigg]\boldsymbol \nu_\alpha \otimes \boldsymbol D_\alpha \bigg), \\
	\mathbb P &= \sum_\alpha  \bigg(\frac{\partial W}{\partial \rho_{\alpha,\alpha}} \boldsymbol \tau_\alpha \otimes \boldsymbol D_\alpha \otimes \boldsymbol D_\alpha + \frac{\partial W}{\partial \vartheta_{\alpha,\alpha}}\rho_\alpha^{-1} \boldsymbol \nu_\alpha \otimes \boldsymbol D_\alpha \otimes \boldsymbol D_\alpha \bigg). \label{eq:doublestress}
\end{align}
The divergence of the $\bb P$ can be computed using \eqref{eq:fiberderivative} and \eqref{eq:tangent_vector_fiber_derivative}:  
\begin{equation}
	\begin{aligned}
		(\mathrm{Div} \,\bbP)_{iA} & = \frac{\partial}{\partial X_B}\sum_{\alpha} \bigg(\bigg[ \frac{\partial W}{\partial \rho_{\alpha,\alpha}} \tau^{(\alpha)}_i + \frac{\partial W}{\partial \vartheta_{\alpha,\alpha}}\rho_\alpha^{-1} \nu^{(\alpha)}_i \bigg] D^{(\alpha)}_A  D^{(\alpha)}_B\bigg) \\
		&= \sum_{\alpha} \bigg(\bigg[ \frac{\partial W}{\partial \rho_{\alpha,\alpha}} \tau^{(\alpha)}_i + \frac{\partial W}{\partial \vartheta_{\alpha,\alpha}}\rho_\alpha^{-1} \nu^{(\alpha)}_i \bigg] \bigg )\!\!\!{\phantom{\Big|}}_{,\alpha} D^{(\alpha)}_A \\
		&= \sum_{\alpha} \bigg[ \bigg ( \frac{\partial W}{\partial \rho_{\alpha, \alpha}} \bigg)\!\!\!{\phantom{\Big|}}_{,\alpha} - \frac{\partial W}{\partial \vartheta_{\alpha,\alpha}}\rho_\alpha^{-1}\vartheta_{\alpha, \alpha}  \bigg] \tau^{(\alpha)}_i D^{(\alpha)}_A\\
		&+ \sum_\alpha \bigg[ \frac{\partial W}{\partial \rho_{\alpha, \alpha}} \vartheta_{\alpha,\alpha} + \bigg(\frac{\partial W}{\partial \vartheta_{\alpha,\alpha}}\bigg)\!\!\!{\phantom{\Big|}}_{,\alpha}
		\rho_\alpha^{-1} - \frac{\partial W}{\partial \vartheta_{\alpha,\alpha}}\rho_\alpha^{-2} \rho_{\alpha,\alpha}   \bigg] \nu^{(\alpha)}_i D^{(\alpha)}_A.
	\end{aligned}
\end{equation}
In direct notation, we have
\begin{align}
	\mathrm{Div}\, \bbP &= \sum_{\alpha} \bigg[ \bigg ( \frac{\partial W}{\partial \rho_{\alpha, \alpha}} \bigg)\!\!\!{\phantom{\Big|}}_{,\alpha} - \frac{\partial W}{\partial \vartheta_{\alpha,\alpha}}\rho_\alpha^{-1}\vartheta_{\alpha, \alpha}  \bigg] \boldsymbol \tau_\alpha \otimes \boldsymbol D_\alpha\\
	&+ \sum_\alpha \bigg[ \frac{\partial W}{\partial \rho_{\alpha, \alpha}} \vartheta_{\alpha,\alpha} + \bigg(\frac{\partial W}{\partial \vartheta_{\alpha,\alpha}}\bigg)\!\!\!{\phantom{\Big|}}_{,\alpha}
	\rho_\alpha^{-1} - \frac{\partial W}{\partial \vartheta_{\alpha,\alpha}}\rho_\alpha^{-2} \rho_{\alpha,\alpha} \bigg] \boldsymbol \nu_\alpha \otimes \boldsymbol D_\alpha,
\end{align}
and thus,
\begin{align}
\bar{\boldsymbol P} = \boldsymbol P - \mathrm{Div}\, \bb P = \sum_{\alpha}  \bigg[ \bigg (\frac{\partial W}{\partial \rho_\alpha} - \bigg ( \frac{\p W}{\p \rho_{\alpha, \alpha}} \bigg )\!\!\!{\phantom{\Big|}}_{,\alpha} \bigg )  \boldsymbol \tau_\alpha \otimes \boldsymbol D_\alpha 
- \rho_\alpha^{-1} \bigg(\frac{\partial W}{\partial \vartheta_{\alpha,\alpha}}\bigg)\!\!\!{\phantom{\Big|}}_{,\alpha}
\boldsymbol \nu_\alpha \otimes \boldsymbol D_\alpha \bigg]. \label{eq:barP}
\end{align}

We now compute 
\begin{align}
	D^2_{\bb F \bb F} W = D_{\bb F} \bb P = \bigg ( \frac{\p}{\p \bb F_{jCD}} \bb P \bigg ) \otimes \boldsymbol e_j \otimes \boldsymbol E_C \otimes \boldsymbol E_D. 
\end{align}
By \eqref{eq:fiber_curvature}, \eqref{eq:fiber_stretch_gradient}, and the chain rule,
\begin{align}
	\frac{\p}{\p \bbF_{jCD}} = \sum_\beta \bigg ( \tau_j^{(\beta)} D_C^{(\beta)} {D_D^{(\beta)}}\frac{\p}{\p \rho_{\beta, \beta}} + \rho_\beta^{-1}  \nu_j^{(\beta)} D_C^{(\beta)} D_D^{(\beta)} \frac{\p}{\p \vartheta_{\beta, \beta}} \bigg),
\end{align}
and thus,
\begin{align}
\begin{split}
	D^2_{\bb F \bb F} W &= \sum_{\alpha, \beta} \frac{\p^2 W}{\p \rho_{\beta, \beta} \p \rho_{\alpha, \alpha}} \boldsymbol \tau_\alpha \otimes \boldsymbol D_\alpha \otimes \boldsymbol D_\alpha \otimes \boldsymbol \tau_\beta \otimes \boldsymbol D_\beta \otimes \boldsymbol D_\beta \\
&+ \sum_{\alpha, \beta} \frac{\p^2 W}{\p \rho_{\beta, \beta} \p \vartheta_{\alpha, \alpha}} \rho_\alpha^{-1} \boldsymbol \nu_\alpha \otimes \boldsymbol D_\alpha \otimes \boldsymbol D_\alpha \otimes \boldsymbol \tau_\beta \otimes \boldsymbol D_\beta \otimes \boldsymbol D_\beta \\
&+ \sum_{\alpha, \beta} \frac{\p^2 W}{\p \vartheta_{\beta, \beta} \p \rho_{\alpha, \alpha}} \rho_\beta^{-1} \boldsymbol \tau_\alpha \otimes \boldsymbol D_\alpha \otimes \boldsymbol D_\alpha \otimes \boldsymbol \nu_\beta \otimes \boldsymbol D_\beta \otimes \boldsymbol D_\beta \\
&+ \sum_{\alpha, \beta} \frac{\p^2 W}{\p \vartheta_{\beta, \beta} \p \vartheta_{\alpha, \alpha}} \rho_\alpha^{-1} \rho_\beta^{-1} \boldsymbol \nu_\alpha \otimes \boldsymbol D_\alpha \otimes \boldsymbol D_\alpha \otimes \boldsymbol \nu_\beta \otimes \boldsymbol D_\beta \otimes \boldsymbol D_\beta. 
\end{split}\label{eq:hessian}
\end{align}

\section{Pantographic microstructures generating incomplete and complete second-gradient continua}

In this section, we give examples of incomplete and complete second-gradient continua that can be synthesized from pantographic microstructures. Throughout, we consider continua containing two initially straight fiber families aligned with the coordinate directions
\begin{align}
	\boldsymbol D_1 = \boldsymbol E_1,
	\qquad
	\boldsymbol D_2 = \boldsymbol E_2.
\end{align}
The stored energies take the form
\begin{align}
	W
	=
	W_0(\boldsymbol F)
	+
	\frac12
	\bbF \cdot \boldsymbol{\mathsf C}(\boldsymbol F)[\bbF],
\end{align}
where \(\boldsymbol{\mathsf C}\) is a sixth-order tensor field possessing the usual major and minor symmetries. For stored energies of this type, the \textit{completeness} of the associated second-gradient continuum is determined entirely by the positivity properties of the quadratic form induced by \(\boldsymbol{\mathsf C}\). More precisely, the continuum is complete if and only if
\begin{align}
	\bbF
	\mapsto
	\bbF \cdot \boldsymbol{\mathsf C}(\boldsymbol F)[\bbF]
\end{align}
is positive definite for every $\boldsymbol F$ in the domain of $W$. Thus, completeness guarantees that a nontrivial second-gradient of the placement contributes positively to the stored energy.

\subsection{Pantographic sheets}
Pantographic sheets are two-dimensional continua whose underlying microstructure consists of families of extensible and continuous fibers endowed with bending stiffness and interconnected by ideal pivots at their intersections. Using a heuristic homogenization procedure \cite{dellIsola2016c}, the stored energy for a pantographic sheet is given by
\begin{align}
	W
	=
	\sum_{\alpha = 1}^2
	\bigg[
	\frac12 k_e(\rho_\alpha - 1)^2
	+
	\frac12 k_b(\vartheta_{\alpha,\alpha})^2
	\bigg].
\end{align}
By \eqref{eq:hessian}, the corresponding Hessian with respect to the second-gradient variable is
\begin{align}
	D^2_{\bbF\bbF}W
	=
	\sum_{\alpha = 1}^2
	k_b\rho_\alpha^{-2}
	\,
	\boldsymbol\nu_\alpha
	\otimes
	\boldsymbol D_\alpha
	\otimes
	\boldsymbol D_\alpha
	\otimes
	\boldsymbol\nu_\alpha
	\otimes
	\boldsymbol D_\alpha
	\otimes
	\boldsymbol D_\alpha.
	\label{eq:D2pantographic_sheet}
\end{align}

We first observe that the quadratic form associated with
\(D^2_{\bbF\bbF}W\)
is nonnegative. Indeed,
\begin{align}
	\bbA
	\cdot
	D^2_{\bbF\bbF}W[\bbA]
	=
	\sum_{\alpha = 1}^2
	k_b\rho_\alpha^{-2}
	\big[
	\bbA
	\cdot
	(
	\boldsymbol\nu_\alpha
	\otimes
	\boldsymbol D_\alpha
	\otimes
	\boldsymbol D_\alpha
	)
	\big]^2
	\geq
	0.
\end{align} 
However, the quadratic form is not positive definite, and hence the continuum is \textit{incomplete}. In particular,
\begin{align}
	D^2_{\bbF\bbF}W
	[
	{\boldsymbol\tau_\gamma}
	\otimes
	\boldsymbol D_\gamma
	\otimes
	\boldsymbol D_\gamma
	]
	=
	\boldsymbol 0,
\end{align}
which shows that second-gradients corresponding to gradients of fiber stretch do not contribute to the energy (see \eqref{eq:hessian}). 

{For this particular model, we now observe corresponding pointwise restrictions on the line double-force and corner-force expressions.} For simplicity, we assume that \(\Omega\) is a rectangle whose sides are parallel to the directions
\[
\boldsymbol E_1 = \boldsymbol D_1,
\qquad
\boldsymbol E_2 = \boldsymbol D_2,
\]
and restrict attention to the right edge \(\partial\Omega_2\), where
\[
\boldsymbol N = \boldsymbol D_1,
\qquad
\boldsymbol T = \boldsymbol D_2.
\]
Along this boundary,
\begin{align}
	\boldsymbol F^{\partial\Omega_2}
	=
	\bar{\boldsymbol P}\boldsymbol N
	=
	k_e(\rho_1-1)\boldsymbol\tau_1
	-
	\rho_1^{-1}k_b\vartheta_{1,11}\boldsymbol\nu_1.
\end{align}
{Since $\{\boldsymbol\tau_1,\boldsymbol\nu_1\}$ forms a basis of the plane, the pointwise constitutive expression places no directional restriction on the line force density. This is a local algebraic statement and does not by itself establish global realization of independently prescribed boundary data.}

The situation is different for line double-forces. Indeed,
\begin{align}
	\boldsymbol D^{\partial\Omega_2}
	=
	k_b\rho_1^{-1}\vartheta_{1,1}\boldsymbol\nu_1,
\end{align}
so that the line double-force is necessarily orthogonal to the current fiber direction \(\boldsymbol\tau_1\). {Thus, the pointwise constitutive expression does not permit a tangential component of the line double-force.}

A similar restriction appears at the corners. In particular, at the corners adjacent to \(\partial\Omega_2\),
\begin{align}
	\boldsymbol F^C
	=
	\boldsymbol 0.
\end{align}
{Thus, the pointwise corner-force expression also vanishes in this aligned rectangular setting.}

\subsection{Bi-pantographic fabric}

We next consider the homogenized continuum obtained in \cite{Barchiesi2020a} that is associated with a discrete microstructure consisting of two orthogonal families of equally spaced pantographic beams connected by hinges at their points of intersection. Unlike the classical pantographic sheet considered previously, the homogenized energy of this microstructure depends not only on the fiber curvatures but also on the fiber stretch gradients. The resulting stored energy takes the form
\begin{equation}\label{eq:strain_energy_bipantograph}
	W
	=
	\sum_{\alpha=1}^2
	\Big[
	f(\rho_\alpha)(\vartheta_{\alpha,\alpha})^2
	+
	g(\rho_\alpha)(\rho_{\alpha,\alpha})^2
	+
	h(\rho_\alpha)
	\Big].
\end{equation}
The functions \(f\), \(g\), and \(h\) depend on the opening angle \(2\gamma \in (0, \frac{\pi}{2})\) between the arms of the discrete pantographic beam and on the material parameters \(k_e\), \(k_f\), and \(k_s\), which characterize the extensional stiffness of the arms, the flexural stiffness of the arms, and the stiffness of the hinges connecting the arms, respectively. Specifically,
\begin{equation}\label{eq:fghequations}
	\begin{aligned}
		f(\rho_\alpha)
		&=
		k_e k_f
		\frac{1-(\rho_\alpha)^2\cos^2\gamma}
		{k_e - (\rho_\alpha)^2\cos^2\gamma
			[k_e-8k_f\cos^2\gamma]},
		\\
		g(\rho_\alpha)
		&=
		k_e k_f
		\frac{(\rho_\alpha)^2\cos^2\!\gamma}
		{
			[1-(\rho_\alpha)^2\cos^2\!\gamma]
			[
			8k_f
			+
			(\rho_\alpha)^2
			(k_e-8k_f\cos^2\!\gamma)
			]
		},
		\\
		h(\rho_\alpha)
		&=
		k_s
		\Big[
		\arccos
		\big(
		1-2(\rho_\alpha)^2\cos^2\!\gamma
		\big)
		-\pi+2\gamma
		\Big]^2,
	\end{aligned}
\end{equation}
In what follows, we require that 
\begin{align}
 k_s \geq 0, \quad k_e, k_f > 0, \qquad (\rho_\alpha)^2 < \frac{1}{\cos^2\gamma}, \quad \alpha = 1, 2.
\end{align}
and thus, 
\begin{align}
	0 \leq h(\rho_\alpha) < \infty, \quad 0 < f(\rho_\alpha), g(\rho_\alpha) < \infty, \quad \alpha = 1, 2.
\end{align}
We remark that this continuum need not be interpreted exclusively as the homogenized limit of a two-dimensional discrete microstructure. From the perspective of synthesis, it is equally natural to view each fiber family as representing a homogenized discrete pantographic beam \cite{Barchiesi2019}.    

By \eqref{eq:hessian}, the Hessian of \(W\) with respect to the second-gradient variable is
\begin{align}
	D^2_{\bbF\bbF}W
	&=
	\sum_{\alpha=1}^2
	2g(\rho_\alpha)
	\boldsymbol\tau_\alpha
	\otimes
	\boldsymbol D_\alpha
	\otimes
	\boldsymbol D_\alpha
	\otimes
	\boldsymbol\tau_\alpha
	\otimes
	\boldsymbol D_\alpha
	\otimes
	\boldsymbol D_\alpha
	\\
	&
	+
	\sum_{\alpha=1}^2
	2f(\rho_\alpha)\rho_\alpha^{-2}
	\boldsymbol\nu_\alpha
	\otimes
	\boldsymbol D_\alpha
	\otimes
	\boldsymbol D_\alpha
	\otimes
	\boldsymbol\nu_\alpha
	\otimes
	\boldsymbol D_\alpha
	\otimes
	\boldsymbol D_\alpha .
\end{align}
Since \(f(\rho_\alpha)>0\) and \(g(\rho_\alpha)>0\), the associated quadratic form is nonnegative. Indeed,
\begin{align}
	\bbA\cdot D^2_{\bbF\bbF}W[\bbA]
	&=
	\sum_{\alpha=1}^2
	2g(\rho_\alpha)
	\big[
	\bbA\cdot
	(
	\boldsymbol\tau_\alpha
	\otimes
	\boldsymbol D_\alpha
	\otimes
	\boldsymbol D_\alpha
	)
	\big]^2
	\\
	&\quad
	+
	\sum_{\alpha=1}^2
	2f(\rho_\alpha)\rho_\alpha^{-2}
	\big[
	\bbA\cdot
	(
	\boldsymbol\nu_\alpha
	\otimes
	\boldsymbol D_\alpha
	\otimes
	\boldsymbol D_\alpha
	)
	\big]^2
	\geq 0 .
\end{align}
However, the quadratic form is still not positive definite on the full space of admissible second gradients. In particular, the quadratic form vanishes when evaluated on mixed second-gradients of the form
\begin{align}
\mathbb A =
	\boldsymbol a
	\otimes
	(
	\boldsymbol D_1\otimes\boldsymbol D_2
	+
	\boldsymbol D_2\otimes\boldsymbol D_1
	),
	\qquad
	\boldsymbol a\in \mathbb R^2.
\end{align}
Consequently, the homogenized bi-pantographic fabric is still incomplete as a second-gradient continuum. Intuitively, the energy detects second derivatives taken twice along each fiber family, but it does not detect mixed second derivatives involving the two fiber directions.

We next examine the corresponding boundary interactions. By \eqref{eq:doublestress} and \eqref{eq:barP},
\begin{align}
	\bbP
	&=
	\sum_{\alpha=1}^2
	\big[
	2g(\rho_\alpha)\rho_{\alpha,\alpha}
	\boldsymbol\tau_\alpha
	+
	2f(\rho_\alpha)\rho_\alpha^{-1}\vartheta_{\alpha,\alpha}
	\boldsymbol\nu_\alpha
	\big]
	\otimes
	\boldsymbol D_\alpha
	\otimes
	\boldsymbol D_\alpha, \\
\bar{\boldsymbol P} &= \sum_{\alpha = 1}^2
\left[
f'(\rho_\alpha)(\vartheta_{\alpha,\alpha})^2
-
g'(\rho_\alpha)(\rho_{\alpha,\alpha})^2
+
h'(\rho_\alpha)
-
2g(\rho_\alpha)\rho_{\alpha,\alpha\alpha}
\right]\boldsymbol\tau_\alpha \otimes \boldsymbol D_\alpha
\\
&\qquad
-
2\rho_\alpha^{-1}
\left[
f'(\rho_\alpha)\rho_{\alpha,\alpha}\vartheta_{\alpha,\alpha}
+
f(\rho_\alpha)\vartheta_{\alpha,\alpha\alpha}
\right]\boldsymbol\nu_\alpha \otimes \boldsymbol D_\alpha.
\end{align}
On the right edge \(\partial\Omega_2\), where
\[
\boldsymbol N=\boldsymbol D_1,
\qquad
\boldsymbol T=\boldsymbol D_2,
\]
we obtain
\begin{align}
	\boldsymbol F^{\partial\Omega_2}
	&=
	\left[
	f'(\rho_1)(\vartheta_{1,1})^2
	-
	g'(\rho_1)(\rho_{1,1})^2
	+
	h'(\rho_1)
	-
	2g(\rho_1)\rho_{1,11}
	\right]\boldsymbol\tau_1
	\\
	&
	-
	2\rho_1^{-1}
	\left[
	f'(\rho_1)\rho_{1,1}\vartheta_{1,1}
	+
	f(\rho_1)\vartheta_{1,11}
	\right]\boldsymbol\nu_1, \\
	\boldsymbol D^{\partial\Omega_2}
	&
	=
	2g(\rho_1)\rho_{1,1}\boldsymbol\tau_1
	+
	2f(\rho_1)\rho_1^{-1}\vartheta_{1,1}\boldsymbol\nu_1 .
\end{align}
{Thus, unlike the pantographic sheet, the pointwise constitutive expressions for the bi-pantographic fabric allow both line forces and line double-forces to have components tangent and normal to the current fiber direction.}

{Nevertheless, the corner-force expression remains restricted in this aligned rectangular setting.} Since \(\bbP\) contains only terms proportional to \(\boldsymbol D_\alpha\otimes\boldsymbol D_\alpha\), we have
\begin{align}
	\bbP[\boldsymbol T\otimes\boldsymbol N]
	=
	\boldsymbol 0
\end{align}
on the edges of a rectangular domain aligned with \(\boldsymbol D_1\) and \(\boldsymbol D_2\). Hence,
\begin{align}
	\boldsymbol F^C
	=
	\boldsymbol 0
\end{align}
at the corners. {Thus, in this aligned rectangular setting, the local constitutive expressions impose no directional restriction on line forces or line double-forces, whereas the corner force vanishes.}

\subsection{Tri-pantographic fabric} Finally, we consider a continuum containing three families of fibers that behave as homogenized pantographic beams \cite{Barchiesi2019}. In addition to the two fiber families aligned with \(\boldsymbol D_1\) and \(\boldsymbol D_2\), we introduce a third initially straight family oriented along
\begin{align}
	\boldsymbol D_3
	=
	\cos\eta\,\boldsymbol D_1
	+
	\sin\eta\,\boldsymbol D_2
	=
	\cos\eta\,\boldsymbol E_1
	+
	\sin\eta\,\boldsymbol E_2,
	\label{eq:D3}
\end{align}
where \(\eta\in(0,\pi)\setminus\{\frac{\pi}{2}\}\). Thus, the third fiber family is not parallel to \(\boldsymbol D_1\) nor \(\boldsymbol D_2\) and introduces a third distinguished direction into the microstructure.

Motivated by the homogenized model for bi-pantographic fabrics, we postulate the stored energy
\begin{equation}\label{eq:strain_energy_tripantograph}
	W
	=
	\sum_{\alpha=1}^3
	\Big[
	f(\rho_\alpha)(\vartheta_{\alpha,\alpha})^2
	+
	g(\rho_\alpha)(\rho_{\alpha,\alpha})^2
	+
	h(\rho_\alpha)
	\Big],
\end{equation}
where the functions \(f\), \(g\), and \(h\) are given by \eqref{eq:fghequations}. Unlike the bi-pantographic model, which arises from homogenization of two-dimensional discrete microstructure \cite{Barchiesi2020a}, the tri-pantographic continuum here is introduced directly at the continuum level. In this interpretation, the microstructure consists of three fiber families, each viewed as representing a homogenized pantographic beam in the sense of \cite{Barchiesi2019}.

{For the tri-pantographic model, we impose the same admissibility conditions as above on all three fiber families; in particular,
\[
	(\rho_\alpha)^2 < \frac{1}{\cos^2\gamma},
	\qquad
	\alpha=1,2,3,
\]
so that \(0<f(\rho_\alpha),g(\rho_\alpha)<\infty\) for \(\alpha=1,2,3\).}

However, the structure of \eqref{eq:strain_energy_tripantograph} suggests a two-dimensional discrete realization. Since the interactions leading to the homogenized bi-pantographic energy \eqref{eq:strain_energy_bipantograph} are uncoupled at the level of the discrete pantographic beams \cite{Barchiesi2020a}, it is natural to conjecture that a two-dimensional discrete microstructure consisting of three families of equally spaced pantographic beams connected by hinges at their intersections would homogenize to an energy of the form \eqref{eq:strain_energy_tripantograph}. Establishing such a homogenization result, however, lies beyond the scope of the present work.

We now show that the \textit{tri-pantographic fabric is complete}. By \eqref{eq:hessian}, the Hessian of \(W\) with respect to the second-gradient variable is
\begin{align}
	D^2_{\bbF\bbF}W
	&=
	\sum_{\alpha=1}^3
	2g(\rho_\alpha)
	\boldsymbol\tau_\alpha
	\otimes
	\boldsymbol D_\alpha
	\otimes
	\boldsymbol D_\alpha
	\otimes
	\boldsymbol\tau_\alpha
	\otimes
	\boldsymbol D_\alpha
	\otimes
	\boldsymbol D_\alpha
	\\
	&
	+
	\sum_{\alpha=1}^3
	2f(\rho_\alpha)\rho_\alpha^{-2}
	\boldsymbol\nu_\alpha
	\otimes
	\boldsymbol D_\alpha
	\otimes
	\boldsymbol D_\alpha
	\otimes
	\boldsymbol\nu_\alpha
	\otimes
	\boldsymbol D_\alpha
	\otimes
	\boldsymbol D_\alpha. \label{eq:quadform}
\end{align}
Since \(f(\rho_\alpha)>0\) and \(g(\rho_\alpha)>0\), the associated quadratic form is nonnegative:
\begin{align}
	\bbA\cdot D^2_{\bbF\bbF}W[\bbA]
	&=
	\sum_{\alpha=1}^3
	2g(\rho_\alpha)
	\Big[
	\bbA\cdot
	(
	\boldsymbol\tau_\alpha
	\otimes
	\boldsymbol D_\alpha
	\otimes
	\boldsymbol D_\alpha
	)
	\Big]^2
	\\
	&
	+
	\sum_{\alpha=1}^3
	2f(\rho_\alpha)\rho_\alpha^{-2}
	\Big[
	\bbA\cdot
	(
	\boldsymbol\nu_\alpha
	\otimes
	\boldsymbol D_\alpha
	\otimes
	\boldsymbol D_\alpha
	)
	\Big]^2
	\geq 0.
\end{align}

To show positive definiteness, suppose that
\begin{align}
	\bbA\cdot D^2_{\bbF\bbF}W[\bbA]=0.
\end{align}
Since all coefficients are strictly positive, every term in the preceding sum must vanish. Moreover, because
\(\{\boldsymbol\tau_\alpha,\boldsymbol\nu_\alpha\}\)
forms a basis of \(\mathbb R^2\) for each $\alpha = 1, 2, 3$, it follows that
\begin{align}
	\bbA
	[
	\boldsymbol D_\alpha
	\otimes
	\boldsymbol D_\alpha
	]
	=
	\boldsymbol 0,
	\qquad
	\alpha=1,2,3.
	\label{eq:Avanishing}
\end{align}
Using
\[
\boldsymbol D_3
=
\cos\eta\,\boldsymbol D_1
+
\sin\eta\,\boldsymbol D_2,
\]
we obtain
\begin{align}
	\boldsymbol 0
	&=
	\bbA
	[
	\boldsymbol D_3
	\otimes
	\boldsymbol D_3
	]
	\\
	&=
	\cos^2\eta\,
	\bbA
	[
	\boldsymbol D_1
	\otimes
	\boldsymbol D_1
	]
	+
	\sin^2\eta\,
	\bbA
	[
	\boldsymbol D_2
	\otimes
	\boldsymbol D_2
	]
	+
	\cos\eta\sin\eta\,
	\bbA
	[
	\boldsymbol D_1
	\otimes
	\boldsymbol D_2
	+
	\boldsymbol D_2
	\otimes
	\boldsymbol D_1
	].
\end{align}
By \eqref{eq:Avanishing} and the assumption
\(
\eta\in(0,\pi)\setminus\{\frac{\pi}{2}\},
\)
we conclude that
\begin{align}
	\bbA
	[
	\boldsymbol D_1
	\otimes
	\boldsymbol D_2
	+
	\boldsymbol D_2
	\otimes
	\boldsymbol D_1
	]
	=
	\boldsymbol 0.
\end{align}
Since
\[
\{
\boldsymbol D_1\otimes\boldsymbol D_1,\,
\boldsymbol D_2\otimes\boldsymbol D_2,\,
\boldsymbol D_1\otimes\boldsymbol D_2
+
\boldsymbol D_2\otimes\boldsymbol D_1
\}
\]
forms a basis for the space of symmetric second-order tensors, it follows that
\[
\bbA[\boldsymbol S]
=
\boldsymbol 0
\]
for every symmetric tensor \(\boldsymbol S\). Since the final two indices of \(\bbA\) are symmetric, we conclude that
\[
\bbA=\boldsymbol 0.
\]
Thus, the quadratic form \eqref{eq:quadform} is positive definite, and the tri-pantographic fabric is complete. 

We remark that the tri-pantographic fabric stored energy \eqref{eq:strain_energy_tripantograph} satisfies
\begin{align}
	W
	&=
	\sum_{\alpha=1}^3
	\Big[
	f(\rho_\alpha)(\vartheta_{\alpha,\alpha})^2
	+
	g(\rho_\alpha)(\rho_{\alpha,\alpha})^2
	+
	h(\rho_\alpha)
	\Big] \\
	&\geq
	\sum_{\alpha=1}^3
	\Big[
	f(\rho_\alpha)(\vartheta_{\alpha,\alpha})^2
	+
	g(\rho_\alpha)(\rho_{\alpha,\alpha})^2
	\Big]
	=
	\frac12
	\bbF
	\cdot
	D^2_{\bbF\bbF}W[\bbF].
\end{align}
{Since the tri-pantographic fabric is complete, its second-gradient quadratic form is positive definite. For each fixed admissible deformation gradient $\boldsymbol F$ and fixed positive moduli $k_e$ and $k_f$, there exists a constant $c=c(\boldsymbol F;k_e,k_f,\gamma,\eta)>0$ such that}
\begin{align}
	W
	\geq
	c
	|\bbF|^2.
\end{align}
{This estimate is pointwise in $\boldsymbol F$. A uniform constant over a class of configurations requires $\boldsymbol F$ to range over a compact subset of the admissible domain that is bounded away from the singular stretch limit.}

{We next examine the pointwise boundary-interaction expressions for a tri-pantographic fabric.}
By \eqref{eq:doublestress} and \eqref{eq:barP},
\begin{align}
	\mathbb P
	&=
	\sum_{\alpha=1}^3
	\big[
	2g(\rho_\alpha)\rho_{\alpha,\alpha}
	\boldsymbol\tau_\alpha
	+
	2f(\rho_\alpha)\rho_\alpha^{-1}\vartheta_{\alpha,\alpha}
	\boldsymbol\nu_\alpha
	\big]
	\otimes
	\boldsymbol D_\alpha
	\otimes
	\boldsymbol D_\alpha, \\
	\bar{\boldsymbol P} &= \sum_{\alpha = 1}^3
	\left[
	f'(\rho_\alpha)(\vartheta_{\alpha,\alpha})^2
	-
	g'(\rho_\alpha)(\rho_{\alpha,\alpha})^2
	+
	h'(\rho_\alpha)
	-
	2g(\rho_\alpha)\rho_{\alpha,\alpha\alpha}
	\right]\boldsymbol\tau_\alpha \otimes \boldsymbol D_\alpha
	\\
	&\qquad
	-
	2\rho_\alpha^{-1}
	\left[
	f'(\rho_\alpha)\rho_{\alpha,\alpha}\vartheta_{\alpha,\alpha}
	+
	f(\rho_\alpha)\vartheta_{\alpha,\alpha\alpha}
	\right]\boldsymbol\nu_\alpha \otimes \boldsymbol D_\alpha.
\end{align}
On the right edge \(\partial\Omega_2\), where
\[
\boldsymbol N=\boldsymbol D_1,
\qquad
\boldsymbol T=\boldsymbol D_2,
\]
we obtain
\begin{align}
	\boldsymbol F^{\partial\Omega_2}
	&=
	\bar{\boldsymbol P}\boldsymbol D_1
	-
	{\frac{d}{dS}\big(\bbP[\boldsymbol D_2\otimes \boldsymbol D_1]\big)}
	\\
	&=
	\left[
	f'(\rho_1)(\vartheta_{1,1})^2
	-
	g'(\rho_1)(\rho_{1,1})^2
	+
	h'(\rho_1)
	-
	2g(\rho_1)\rho_{1,11}
	\right]\boldsymbol\tau_1
	\\
	&\quad
	-
	2\rho_1^{-1}
	\left[
	f'(\rho_1)\rho_{1,1}\vartheta_{1,1}
	+
	f(\rho_1)\vartheta_{1,11}
	\right]\boldsymbol\nu_1
	\\
	&\quad
	+
	\cos\eta
	\left[
	f'(\rho_3)(\vartheta_{3,3})^2
	-
	g'(\rho_3)(\rho_{3,3})^2
	+
	h'(\rho_3)
	-
	2g(\rho_3)\rho_{3,33}
	\right]\boldsymbol\tau_3
	\\
	&\quad
	-
	2\rho_3^{-1}\cos\eta
	\left[
	f'(\rho_3)\rho_{3,3}\vartheta_{3,3}
	+
	f(\rho_3)\vartheta_{3,33}
	\right]\boldsymbol\nu_3
	\\
	&\quad
	-
	\cos\eta\sin\eta
	\left[
	2g(\rho_3)\rho_{3,3}\boldsymbol\tau_3
	+
	2f(\rho_3)\rho_3^{-1}\vartheta_{3,3}\boldsymbol\nu_3
	\right]_{,2}, \\
{\boldsymbol D^{\partial \Omega_2}} &=
	2g(\rho_1)\rho_{1,1}
\boldsymbol\tau_1
+
2f(\rho_1)\rho_1^{-1}\vartheta_{1,1}
\boldsymbol\nu_1 \\
&+ \cos^2 \eta\, 2g(\rho_3)\rho_{3,3}
\boldsymbol\tau_3
+
2 \cos^2 \eta \, f(\rho_3)\rho_3^{-1}\vartheta_{3,3}
\boldsymbol\nu_3.
\end{align}
{Thus, on the right edge $\partial\Omega_2$, the pointwise line-force and line-double-force expressions for the tri-pantographic fabric may have both tangential and normal components.} In contrast with the bi-pantographic fabric, the third fiber family also produces nonzero mixed edge interactions through \(\bbP[\boldsymbol T\otimes\boldsymbol N]\), which are absent when only the two coordinate-aligned fiber families are present.

At the upper-right corner $\boldsymbol X_{\mathrm{top}} \in \partial\Omega_2$, the adjacent boundary segments are characterized by
\begin{align}
\boldsymbol N=\boldsymbol D_1,
\qquad
\boldsymbol T=\boldsymbol D_2
\end{align}
along the right edge and
\begin{align}
\boldsymbol N=\boldsymbol D_2,
\qquad
\boldsymbol T=-\boldsymbol D_1
\end{align}
along the top edge. Consequently,
\begin{align}
	\big(\bbP[\boldsymbol T\otimes\boldsymbol N]\big)^+
	&=
	-\cos\eta\sin\eta
	\Big[
	2g(\rho_3)\rho_{3,3}\boldsymbol\tau_3
	+
	2f(\rho_3)\rho_3^{-1}\vartheta_{3,3}\boldsymbol\nu_3
	\Big],
	\\
	\big(\bbP[\boldsymbol T\otimes\boldsymbol N]\big)^-
	&=
	\cos\eta\sin\eta
	\Big[
	2g(\rho_3)\rho_{3,3}\boldsymbol\tau_3
	+
	2f(\rho_3)\rho_3^{-1}\vartheta_{3,3}\boldsymbol\nu_3
	\Big].
\end{align}
Hence,
\begin{align}
	\boldsymbol F^{\boldsymbol X_{\mathrm{top}}}
	=
	\big(\bbP[\boldsymbol T\otimes\boldsymbol N]\big)^-
	-
	\big(\bbP[\boldsymbol T\otimes\boldsymbol N]\big)^+
	=
	2\cos\eta\sin\eta
	\Big[
	2g(\rho_3)\rho_{3,3}\boldsymbol\tau_3
	+
	2f(\rho_3)\rho_3^{-1}\vartheta_{3,3}\boldsymbol\nu_3
	\Big].
\end{align}
Similarly, at the lower-right corner $\boldsymbol X_{\mathrm{bot}} \in \partial\Omega_2$, the adjacent boundary segments are characterized by
\begin{align}
\boldsymbol N=\boldsymbol D_1,
\qquad
\boldsymbol T=\boldsymbol D_2
\end{align}
along the right edge and
\begin{align}
\boldsymbol N=-\boldsymbol D_2,
\qquad
\boldsymbol T=\boldsymbol D_1
\end{align}
along the bottom edge. A direct calculation yields
\begin{align}
	\boldsymbol F^{\boldsymbol X_{\mathrm{bot}}}
	=
	-2\cos\eta\sin\eta
	\Big[
	2g(\rho_3)\rho_{3,3}\boldsymbol\tau_3
	+
	2f(\rho_3)\rho_3^{-1}\vartheta_{3,3}\boldsymbol\nu_3
	\Big].
\end{align}
{In particular, unlike the pantographic sheet and the bi-pantographic fabric, the pointwise corner-force expression for the tri-pantographic fabric spans the plane as $\rho_{3,3}$ and $\vartheta_{3,3}$ vary.}

\section{Conclusion}
In this work, we proposed a definition of \textit{completeness} for two-dimensional second-gradient elastic continua that requires local positive definiteness of the Hessian of the stored energy with respect to the second-gradient variable. {\textit{Intuitively}, this condition gives strictly positive quadratic control of every admissible second-gradient increment.}

The notion of completeness was then studied through a broad class of fibrous second-gradient continua with pantographic microstructures, whose stored energies depend on fiber stretch, stretch gradients, and curvature. {Classical pantographic sheets were shown to be incomplete; bi-pantographic fabrics were shown to enlarge the class of second-gradient components detected by the energy while remaining incomplete; and the introduction of a third fiber family was shown to yield a complete continuum.}  We also propose a corresponding discrete three-family architecture for the complete continuum, while leaving its rigorous homogenization open. These examples illustrate how the architecture of the underlying microstructure influences the completeness properties of the resulting continuum model.

The examples studied here also suggest a connection between completeness and the class of boundary interactions that a continuum can support. At the heuristic level, this connection is already visible in the expressions for the line force, line double-force, and corner force in terms of the double-stress tensor
\begin{align}
	\mathbb P
	=
	\frac{\partial W}{\partial \mathbb F_{iAB}}
	\boldsymbol e_i
	\otimes
	\boldsymbol E_A
	\otimes
	\boldsymbol E_B.
\end{align}
If the stored energy does not depend on certain components of the second gradient, then the corresponding components of the double-stress are absent. Through \eqref{eq:boundary_conditions_edges_2nd_gradient} and \eqref{eq:bc_edges_Lagrangian}, this may lead to restrictions on the line forces, line double-forces, and corner forces that the continuum can support.

{A related question concerns the linearized theory. At a fixed configuration, completeness makes the principal second-gradient quadratic form pointwise positive definite. Coercivity of the full weak bilinear form on $H^2(\Omega;\mathbb R^2)$, however, would additionally require a uniform lower bound over $\Omega$, bounded coefficients, and suitable control of the first-gradient and coupling terms. Because the Hessian seminorm has an affine kernel, one must also impose essential boundary conditions or normalization conditions that eliminate this kernel; prescribing displacement alone on an arbitrary nontrivial boundary portion is not sufficient without further geometric hypotheses. Under such assumptions, existence and uniqueness may then be sought through the Lax-Milgram theorem \cite{Evans2010}. Purely natural problems would additionally require appropriate equilibration conditions on the loads, whereas mixed problems with nonempty essential boundary data must be treated separately. Establishing precise mathematical relationships between completeness, uniform coercivity, well-posedness, and admissible boundary interactions remains an open direction for future research.}

\bibliographystyle{abbrv}
\bibliography{literature}

\bigskip

\footnotesize

\noindent \textsc{Department of Mathematics, The University of North Carolina at Chapel Hill, USA}\\
\noindent \textit{E-mail address}: \texttt{crodrig@email.unc.edu}

\bigskip

\noindent \textsc{Dipartimento di Architettura, Design e Urbanistica, Universit\`a di Sassari, Italy}\\
\noindent \textit{E-mail address}: \texttt{ebarchiesi@uniss.it}

\bigskip

\noindent \textsc{Department of Mechanical Engineering, Eindhoven University of Technology, The Netherlands}\\
\noindent \textit{E-mail address}: \texttt{s.r.eugster@tue.nl}

\bigskip

\noindent \textsc{Dipartimento di Ingegneria Civile, Edile-Architettura e Ambientale, Universit\`a dell'Aquila, Italy}\\
\noindent \textit{E-mail address}: \texttt{ivan.giorgio@univaq.it}

\bigskip

\noindent \textsc{Dipartimento di Ingegneria Civile, Edile-Architettura e Ambientale, Universit\`a dell'Aquila, Italy}\\
\noindent \textsc{CNRS Fellow, ENS Paris-Saclay, France}\\
\noindent \textit{E-mail address}: \texttt{francesco.dellisola@univaq.it}

\end{document}